\documentclass[onecollarge,natbib]{svjour2}
\bibpunct{[}{]}{;}{n}{}{,} 
\smartqed  
\usepackage{graphicx}
\def\be{\begin{eqnarray} &&} 
\def\nonu{\nonumber \\ &&} 
\def\ee{\end{eqnarray}} 
\newcommand{\bm}[1] {\mbox{\boldmath{$#1$}}}
\newcommand{\blf}[1]{\bf  {\tilde #1}}
\def\sumint{\int \! \!\ \! \! \! \! \!\ \! \! \!\! \!\sum}

\journalname{Few-Body Systems}
\begin{document}

\title{Light-Front Dynamics and the $^3He$ Spectral Function
}


\author{Emanuele Pace         \and
        Alessio Del Dotto   \and
       Leonid Kaptari   \and
        Matteo Rinaldi    \and
        Giovanni Salm\'e      \and
        Sergio Scopetta       \and
}


\institute{E. Pace \at
      Dipartimento di Fisica,  Universit\`a di Roma ÒTor VergataÓ and INFN, Sezione di Roma Tor Vergata, Italy \\
              Tel.: +39-0672594565\\
              Fax: +39-062023507\\
              \email{pace@roma2.infn.it}           
           \and
           A. Del Dotto  \at
       INFN, Sezione di Roma, Italy
               \and
           L. Kaptari  \at
         Bogoliubov Lab. Theor. Phys., 141980, JINR, Dubna, Russia   \and
           M. Rinaldi  \at
      Dipartimento di Fisica e Geologia,    Universit\`a di Perugia and INFN, Sezione di Perugia, Italy
            \and
           G. Salm\'e \at
          INFN, Sezione di Roma, Italy  \and
          S. Scopetta  \at
       Dipartimento di Fisica e Geologia,   Universit\`a di Perugia and INFN, Sezione di Perugia, Italy
}

\date{Received: date / Accepted: date}

\maketitle

\begin{abstract}
Two topics are presented. The first one is a novel approach for a Poincar\'e covariant description of nuclear dynamics  based on 
light-front Hamiltonian dynamics. 
The key quantity is the light-front spectral function, where both 
normalization and momentum sum rule can be satisfied at the same time. 
Preliminary results are discussed for an initial analysis of the role of relativity in the EMC effect in $^3He$.
A second issue, very challenging, is considered in a non-relativistic framework, namely a distorted spin-dependent spectral function for $^3He$ in order to take care of the final state interaction between the observed pion and the remnant in  
semi-inclusive deep inelastic electron scattering off polarized $^3He$.
 The generalization of the analysis  within the light-front dynamics is outlined. 
\end{abstract}

\section{Introduction}
\label{intro}

The standard model of few-nucleon systems, where nucleon and pion degrees of freedom are taken into account, has achieved a very high degree of sophistication.
Nonetheless, one should try to fulfill, as much as possible, the relativistic constraints, dictated by the covariance with respect to the Poincar\'e group, if processes involving nucleons with large 3-momentum are considered and a high precision is needed. This is the case if one studies, e.g., i) the nucleon structure functions (unpolarized and polarized); ii)  semi-inclusive deep inelastic (SIDIS) processes, iii) signatures of short-range correlations.
At least, one should carefully deal with the boosts of the nuclear states, $|\Psi_{init}\rangle$ and $|\Psi_{fin}\rangle$ !

 In particular a relativistic treatment is important to accurately describe the JLab program @ 12 GeV for few-body systems (see, e.g., \cite{DIS}, \cite{SIDIS}, \cite{06010}).
 
Our key tool for a relativistic description of few-body nuclei is a Poincar\'e covariant spectral function (SF) built up within the light-front Hamiltonian dynamics  (LFHD).

Indeed, the Relativistic Hamiltonian Dynamics (RHD) of an interacting system with a fixed number of on-mass-shell constituents (see, e.g., \cite{KP}) 
plus the Bakamijan-Thomas construction of the Poincar\'e generators \cite{Baka}  allow one to give a  fully Poincar\'e covariant description of DIS, SIDIS and deeply virtual Compton scattering (DVCS) off $^3He$. The light-front (LF) form of RHD is adopted \cite{KP}, which has 
a subgroup structure of the LF boosts (with a separation of the intrinsic motion from the global one: very important for us!) and a meaningful Fock expansion (see, e.g., \cite{Brodsky}).
Furthermore, within the LFHD one can take advantage of the whole successful non-relativistic (NR) phenomenology 
that has been 
developed for the nuclear interaction.

In Section 2 the LF spin-dependent (SD) SF obtained from the LF wave functions for the two- and the three-nucleon systems is described. In Section 3 the LF SF is applied to study the role of relativity for the EMC effect in $^3He$ and preliminary results are presented. In Section 4  the actual possibility to get information on the neutron structure from SIDIS experiments on 
$^3He$  is shown within a non-relativistic approach  which include the final state interaction (FSI) through a distorted SF \cite{Kaptari,DelDotto,Kaptari1}. Preliminary results with the LF SF  are reported. 
In Section 5 conclusions and perspectives  are drawn.

\section{Light-Front Dynamics and the Light-Front Spectral Function}
An explicit construction of the 10 Poincar\'e generators that fulfills the proper commutation rules in presence of interactions was given by Bakamjian and Thomas \cite{Baka} through
 the mass operator, M : i) only  M contains the interaction; ii) it generates the dependence upon the interaction of the three dynamical generators in LFHD, namely $P^-$ and the  transverse rotations $\vec F_\perp$.
M
is obtained adding to the free mass $M_0$ of the system an interaction $V$. There are two possibilities: $M^2 = M_0^2 + U$ (then for  two particles one can easily embed the NR phenomenology) or $M = M_0 + V$.
The interaction, $U$ or $V$, must commute with all the kinematical generators, and with the non interacting spin. Then it has to be invariant for translations, as in the NR case, and the angular momentum is a conserved quantity.

The full theory must fulfill the macroscopic locality. This property could be implemented by using interaction-dependent, unitary operators: the packing operators (see, e.g., \cite{KP}). Their effects should be small, and therefore they will be neglected in what follows, but in principle should be investigated.

For the three-body case the mass operator is
$M_{BT}(123)= {{M_0(123)}}+ V^{BT}$,
where
$M_0(123)= \sqrt{m^2 +k^2_1}+\sqrt{m^2 +k^2_2}+\sqrt{m^2 +k^2_3}$
is the free mass operator,	$V^{BT}$	a Bakamjian-Thomas (BT) two-body and  three-body force, and $k_i~(i=1,2,3)$ are intrinsic momenta with	
${\bf k}_1 +{\bf  k}_2 +{\bf k}_3=0$
 \cite{KP}.

The NR mass operator is written as
$M^{NR}=3m + \sum_{i=1,3} {k^2_i / 2m}
+V^{NR}_{12}+V^{NR}_{23}+V^{NR}_{31}+V^{NR}_{123}$
and must obey  the commutation rules proper of the Galilean group, leading to translational and rotational invariance.
 Those properties are analogous to the ones in the BT construction. This allows us to consider the standard NR mass operator as a sensible BT mass operator, and embed it in a Poincar\'e covariant approach:
$M_{BT}(123) \sim M^{NR}~$.

Coupling  angular momenta is 
accomplished within the {{Instant Form (IF) of RHD}} through the 
{{Clebsch-Gordan coefficients}}. 
 {{To embed this machinery in the LFHD one needs  the
{{Melosh rotations}} that relate  LF and 
IF spin wave functions. For a particle of LF momentum $\tilde k\equiv\{k^+,\vec
k_\perp\}$}} one has
\be
|{\blf k};s \sigma \rangle_{LF} =
~\sum_{ \mu} ~
 D^{s}_{\mu\sigma }
\left[{\cal R}^\dagger_M ({\blf k})\right] ~|{\bf k}; s \mu\rangle_{IF}
\ee
where
{{ $D^{s}_{\sigma,\sigma'}(R^\dagger_{M}(\blf k))$}} 
is the standard Wigner function
and {{ $R_{M}(\blf k)$}} is the rotation between the rest frames of the 
 particle reached through 
a LF boost or a canonical, rotationless boost.

Then for  quantities, like  the spin-dependent SF,  taking care of the Melosh rotations, one obtains
{{$$ O^{LF}_{\sigma',\sigma} =\sum_{\bar \sigma',\bar \sigma}~
D^{1/2}_{\sigma',\bar \sigma'}(R_{M})~O_{\bar \sigma',\bar \sigma}^{IF}~
D^{1/2}_{\bar \sigma, \sigma}(R^\dagger_{M})$$}}
The NR
spin-dependent SF 
for a nucleus of mass number A is defined 
through 
its matrix
elements
\vspace{-0.1cm}
\be 
 P_{\sigma,\sigma',\cal{M}}^{\tau} ({\vec{p},E})=
 \sum\nolimits\limits_{{f}_{(A-1)}}
~\langle{\vec{p},\sigma \tau;\psi
}_{f_{(A-1)}} |{\psi }_{\cal{J}\cal{M}}\rangle~\langle{\psi
}_{\cal{J}\cal{M}}|{\psi }_{f_{(A-1)}};\vec{p},\sigma ' \tau \rangle ~  
\delta (E-{E}_{f_{(A-1)}}+{E}_{A})
\label{spec0}
\ee
where $|{\psi}_{\cal{J}\cal{M}}\rangle$ is the ground state of the nucleus with energy $E_A$ and polarized along
$\vec{S}$, $|{\psi }_{f_{(A-1)}}\rangle$  an eigenstate of the (A-1)-nucleon
system with energy $E_{f_{(A-1)}}$, interacting with the same interaction of the nucleus,
$|\vec{p},\sigma \tau\rangle$  the plane wave for the nucleon $\tau = \pm 1/2$, with momentum 
$\vec{p}$ in the rest frame and  spin
along the z-axis equal to $\sigma$ \cite{cps,cps1}. 
 
 To obtain  within the LFHD a Poincar\'e-covariant 
 spin-dependent 
 SF for a three-particle system in the bound state $|\Psi_{0};S, T_z \rangle$, polarized along
$\vec{S}$, 
 let us replace the NR overlaps 
$\langle{\vec{p},\sigma \tau;\psi}_{f_{(A-1)}} |{\psi }_{\cal{J}\cal{M}}\rangle$
with their LF counterparts 
$_{LF}\langle \tau_{S},T_{S};\alpha,\epsilon ;J_{z}J;\tau\sigma,\tilde{\bm \kappa}|\Psi_{0}; S, T_z\rangle$,
that depend upon the energy $\epsilon$ of the  system of two fully interacting particles (say, particles 2 and 3) and upon the intrinsic momentum, 
$\tilde{\bm \kappa}$, of the third particle (say, particle 1) in the intrinsic reference frame of the cluster (1,23).
Then, the LF spin-dependent 
SF for 
 the three-nucleon system ($^3He$ or $^3H$) 
   is \cite{DPSS}
\be
 \hspace{-0.4cm}{{{\cal {P}}^{\tau}_{\sigma'\sigma}(\xi,{\bm \kappa}_\perp,\kappa^-,S)}
= 
\left|{\partial \kappa^+\over \partial \xi}\right|
~\sumint  d\epsilon~\rho(\epsilon) ~
\delta\left( \kappa^- -M_3+{M^2_S +|{\bm \kappa}_\perp|^2 \over (1-\xi)M_3}
\right)} 
~
~\times
\nonu
\hspace{-0.4cm} 
\sum_{J J_{z}\alpha}\sum_{T_{S}\tau_{S} } ~
_{LF}\langle  \tau_{S},T_{S} , 
\alpha,\epsilon; J J_{z}; \tau\sigma',\tilde{\bm \kappa}|\Psi_{0}; S,T_z
\rangle
  ~\langle S,T_z;
\Psi_0|\tilde{\bm \kappa},\sigma\tau; J J_{z}; 
\epsilon, \alpha, T_{S}, \tau_{S}\rangle_{LF} 
\label{LFspf}
\ee
where $\tau= \pm 1 /2$, $M_3$ is  the nucleus mass, 
$\epsilon$  the intrinsic energy of the
 two-nucleon eigenstate, $\rho(\epsilon)$  the corresponding density of states  
($\rho(\epsilon)  =  \sqrt{\epsilon ~ m} ~ m/2$ for the two-body continuum states and
$\rho(\epsilon)  = 1$ for the deuteron  bound state),
 $J$
  the spin, $T_{S}$
  the isospin  of the two-body state, 
 $\alpha$ the set of quantum numbers needed to completely specify this
  eigenstate,  and
  { $M_S=2\sqrt{m^2 +m\epsilon}$}  its mass. From ~$\xi, M_S, {\bm \kappa}_\perp$~ one can define
~{$\kappa^+=\xi{\cal M}_{0}(1,23)$,  where ${\cal M}_0(1,23)$ is the free mass of the cluster (1,23)
\be
 {\cal M}^2_{0}(1,23)={m^2 +|{\bm \kappa}_\perp|^2 \over \xi}+
{M^2_S +|{\bm \kappa}_\perp|^2 \over (1-\xi)}
\ee
  
The overlap 
$_{LF}\langle \tau_{S},T_{S};\alpha,\epsilon ;J_{z}J;\tau\sigma,\tilde{{\bm \kappa}}|\Psi_{0}; S, T_z\rangle$ is  defined as follows \cite{DPSS}
\be
_{{LF}}\langle  \tau_{S},T_{S} , 
\alpha,\epsilon; J_{z} J; \tau\sigma,{{\tilde{\bm \kappa}}}|\Psi_{0}; S, T_z
 \rangle = \sum_{\sigma'_1}~ D^{{1 \over 2}} [{\cal R}_M^\dagger 
(\tilde{\bm \kappa} )]_{\sigma\sigma'_1}
\sum_{\tau_2,\tau_3} \int d{\bf  k}_{23} 
~\sum_{\sigma_2,\sigma_3}
~\sqrt{ 
  {\kappa^+ E_{23} \over k^+ E_S}}~\times \nonu
\sqrt{(2 \pi)^3~k^+ {\partial k_z\over 
  \partial k^+}}
~~ _{{ {IF}}}\langle  \tau_S, T_{S},
\alpha,\epsilon; J_{z} J |{\bf k}_{23};\sigma_2,\sigma_3;\tau_2,\tau_3
\rangle 
\langle \tau_3,\tau_2,\tau; \sigma_3, \sigma_2, \sigma'_1;
{{\bf k}},{\bf k}_{23}| \Psi_{0}; S,T_z \rangle_{{{IF}}} 
\label{overl}
 \ee
where ${\bf k}_{23}$ is the intrinsic momentum of the (23) pair, ${\bf k}$ is the intrinsic nucleon momentum in the (123) system
 ({${\bf k}_\perp={\bm \kappa}_\perp$, since we choose the $^3He$
 transverse momentum ${\bf P}_\perp=0$}),
  {$k^+ = \xi~ M_0(123)= 
 \kappa^+ ~M_0(123)/ {\cal M}_0(1,23)$, with $M_0(123)$ the free mass of the three-particle system
  \be
  ~~~~~~~M^2_0(123)={m^2 +|{\bm k}_\perp|^2 \over \xi}+{M^2_{23} +
  |{\bm k}_\perp|^2 \over (1-\xi)}
  \ee
  and $~M^2_{23}= 4 (m^2 +|{\bf k}_{23}|^2)$
  the mass of the spectator pair without interaction ! 
In Eq. (\ref{overl}) one has  { $k_z= { 1\over 2} ~\left[k^+ -{(m^2+|{\bm \kappa}_\perp|^2 ) / k^+}
 \right]$, $E_{23}=\sqrt{M^2_{23}+|{\bf k}|^2}$ and $E_S=\sqrt{M^2_S+|{\bm
 \kappa}|^2}$}.

In the actual calculations, we identified the IF overlaps of Eq. (\ref{overl}) with the NR wave functions for the two-nucleon and the three-nucleon \cite{pisa} systems,  corresponding to the NN interaction AV18 \cite{AV18}. 

\section{Light-front momentum distribution and
preliminary results for the EMC effect}
  From the LF  SF one can obtain the light-cone momentum distribution {{$f^A_{p(n)}(z)$}} 
 \be
 f^A_{\tau}(z)  = \int_0^1 d\xi \int d  {\bm \kappa}_\perp\int d\kappa^- ~{1 \over 2 (2 \pi)^3 \kappa^+} ~
 Tr \left[{{{\cal P}^{\tau}(\xi,{\bf k}_\perp, \kappa^-,S)}}\right]~ 
 \delta\left(z - {\xi M_A\over m} \right )
 \ee
that fulfills the following relations, given the SF normalization 
\vspace{-1mm}
\be
{{\int_0^{M_A/m} dz~f^A_{\tau}(z)=1 }} \quad \quad \quad
{{
{N}}_A= {1 \over A}\int_0^{M_A/m} dz~
\left [Z f^A_{p}(z) + (A-Z) f^A_{n}(z)\right] =1} 
\ee
\vspace{-1mm}
The symmetry of the three-body bound state naturally entails the momentum sum rule (see \cite{DPSS})
\be
{{MSR}={1 \over A}\int_0^{M_A/m} dz~z~\left [Z f^A_{p}(z) + (A-Z) f^A_{n}(z)\right]
 ={ 1\over A} }
 \ee
 We evaluated the nuclear structure function { {$ F^A_2(x)$}} ($x=Q^2/2m\nu$)  as a convolution of the nuclear spectral function and of the structure function of the nucleon
 \be
\hspace{-7mm} F_2^A(x)  =
\int_0^1 {d \xi}
\int d{\bm \kappa}_\perp
\int 
\frac{d  \kappa^- }{2(2 \pi)^3  \kappa^+} 
\left [ Z Tr \, {\mathcal{P}}^p(\xi,{\bm \kappa}_\perp, \kappa^- ) 
F_2^p\left({x \over \xi}
\right) + N Tr \, {\mathcal{P}}^n(\xi,{\bm \kappa}_\perp, \kappa^- )  
F_2^n\left({x \over \xi}
\right ) \right ]
\end{eqnarray}

\vspace{1mm} 
Then, to investigate whether the LF SF can affect the EMC effect, the ratios
\be
{{
R^A_2(x)={~ F^A_2(x)\over Z~F^p_2(x)+(A-Z)~F^n_2(x)}}}
\ee
and  $R^{He}_2(x)/R^D_2(x)$ were evaluated.   The Pisa group wave function  \cite{pisa}, 
corresponding to the NN interaction AV18 \cite{AV18}, was used.
For the two-body channel an exact calculation was performed. In the three-body channel  average values for $k_{23}$ were inserted in Eq. (\ref{overl}), different for the proton and the neutron (113.53 MeV  and  91.27 MeV, respectively), corresponding to the average kinetic energy of the intrinsic motion of the spectator pair in the continuum part of the spectrum. We checked in the two-body channel that using  the average value $<k_{23}>$= 136.37 MeV, proper for this channel,  a result  similar  to the exact calculation is obtained. Our preliminary results are shown in Fig. 1 and encourage us in performing the full LF calculation. 

With the same approximation used for the  three-body channel, 
 we  checked that} 
{${{MSR_{calc}= 0.3331}}$}.
\begin{figure}
\centering
 \includegraphics[width=11.cm]{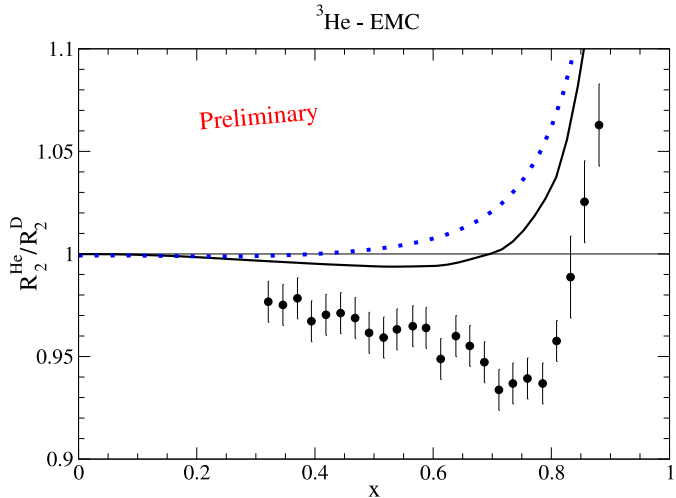}
\caption{EMC effect for $^3He$. Solid line: result for the LF SF, with the exact calculation for the 2-body channel, and an average energy in the 3-body contribution: $<k_{23} >$= 113.53 MeV (proton), $< k_{23} >$= 91.27 MeV (neutron).
Dotted line: result 
with the approach of Ref. \cite{Sauer} for the SF. Experimental data are from Ref. \cite{Seely}.}
\label{fig1}       
\end{figure}
\vspace{-1mm}
\section{Extraction of neutron asymmetries from SIDIS experiments off $^3He$}
In Ref. \cite{mio} it was shown, within the plane wave impulse approximation (no interaction, after the electron scattering off one of the $^3He$ nucleons, between the measured fast $\pi$, the remnant debris and the interacting two-nucleon recoiling system) and using the NR SF of 
Ref. \cite{cps},
 that the formula \cite{neutr}
\be
  A_n \simeq {1 \over 
{{p_n}} f_n} \left 
( {{A^{exp}_3}} - 2 
{{p_p}} f_p
{{A^{exp}_p}} \right )~, \quad 
\label{extrac}
\ee
already widely used to extract neutron asymmetries in DIS from experiments on $^3He$, works also in SIDIS, both for the Collins and Sivers single spin asymmetries. Nuclear effects are hidden in the effective polarizations (EP) $p_p=-0.023$ and $p_n=0.878$ and in the dilution factors, $f_{p(n)}$.

To investigate whether the formula (\ref{extrac}) can be safely applied even in presence of the FSI, in Ref. \cite{Kaptari1} a generalized eikonal approximation (GEA) is used to take care of the FSI through
a distorted spin-dependent SF in a NR approach.
The relevant part of the ({{GEA-distorted}}) spin-dependent
SF is
\vskip 2mm
$
\quad \quad \quad
{\cal P}_{||}^{{{FSI}}}=
{\cal O}^{{FSI}}_{\frac12 \frac12}-
{\cal O}^{{{FSI}}}_{-\frac12 -\frac12}
\quad \quad \quad \quad
$
\quad \quad  \quad \quad ${\cal O}^
{{{FSI}}}_{\lambda\lambda'}(p_N,E) ~{=}$
\vskip -3mm
\begin{eqnarray}&& \!\!\!\!\!\!\!\! 
\hspace{-0.2cm} = \sum \! \!\! \!\! \!\!\int_{~\epsilon^*_{A-1}} \hspace{-4mm}\rho\left(
\epsilon^*_{A-1}\right) \langle   
S_A, 
{{\bf P_A}}
| ({{\hat S_{Gl}}})
\{\Phi_{\epsilon^*_{A-1}},\lambda',{\bf p}_N\} \rangle~
 \langle  
({{\hat S_{Gl}}})
\{\Phi_{\epsilon^*_{A-1}},
\lambda,{\bf p}_N\}|  S_A,{{\bf P_A}}\rangle
\delta\left( E- B_A-\epsilon^*_{A-1}\right)
\end{eqnarray}
where 
$\, {{\hat S_{Gl}}} 
({\bf r}_1,{\bf r}_2,{\bf r}_3)=
\prod_{i=2,3}\bigl[1-\theta(z_i-z_1)
{{\Gamma}}({\bf b}_1-{\bf b}_i,{ z}_1-{z}_i)
\bigr]
$ 
is a {Glauber}} operator  \cite{Kope} which takes care of  hadronization and FSI. The model of Ref. \cite{Kope1} for the (generalized) {{profile function} ${{\Gamma({{\bf b}},z)}}$,
lready successfully applied to $^2H(e,e'p)X$ \cite{Ciofi}, is adopted.


\begin{figure}[h]
\includegraphics[width=14.cm]{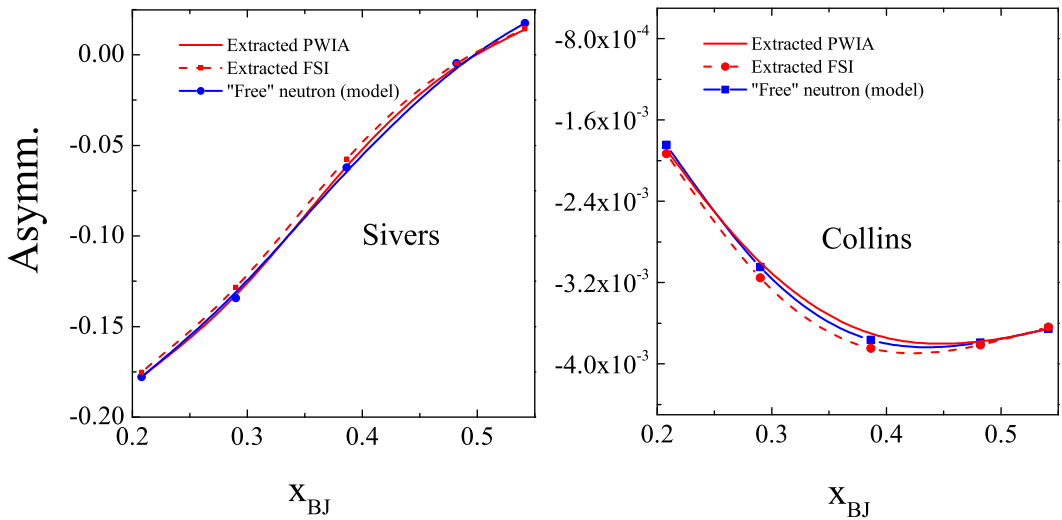}
\vskip 0.1cm
\vskip -1.mm
\caption{Neutron asymmetries extracted through Eq. (\ref{extrac}), from the Sivers (left panel) and Collins (right panel) asymmetries, with and without FSI, in the actual kinematics of JLab \cite{SIDIS}. Preliminary results to appear in \cite{Kaptari1}.}
\label{asymm}
\end{figure}
{{It occurs that effects of GEA-FSI in the dilution factors and in the
EP compensate each other
to a large extent: ${p_p^{FSI}} f_p^{FSI} \approx {p_p} f_p$, ${p_n^{FSI}} f_n^{FSI} \approx {p_n} f_n$ \cite{DelDotto}. Then the {{usual extraction}} of Eq. (\ref{extrac}) is safe, as shown at $E_i=$ 8.8 GeV in Fig. \ref{asymm}.
}}

Relativity effects  seem weak, since LF longitudinal and transverse EPs change little from the ones obtained with the NR SF \cite{Pace}. 
Then the usual extraction procedure should work well also within the LF approach. Concerning FSI, we plan to include in our LF description the FSI between the jet produced from the hadronizing quark and the two-nucleon spectator system through the GEA of \cite{Ciofi}.

\section{Conclusions and Perspectives}
A Poincar\'e covariant description of  A=3 nuclei, based on the LF Hamiltonian dynamics, has been proposed. The Bakamjian-Thomas construction of the Poincar\'e generators allows one to embed the successful NR phenomenology for few-nucleon systems in a Poincar\'e covariant framework. A LF SF can be defined that exactly fulfills both the normalization and the momentum sum rule.
The nucleon SF for $^3He$, has been evaluated by approximating the IF overlaps in Eq. (\ref{overl}) with their NR counterparts, calculated with the AV18 NN interaction, since fulfills rotational and translational symmetries.

A first test of our approach is the EMC effect for $^3He$. The 2-body contribution to the nucleon SF has been calculated with the full expression, while for the 3-body contribution an average value for  $<k_{23}^2>$ has been used. In the comparison with experimental data encouraging improvements clearly appear  with respect to the non-relativistic result.
The next step will be the  full calculation of the EMC effect for $^3He$, including the exact 3-body contribution.

An investigation of SIDIS processes off $^3He$ beyond the NR, impulse approximation (IA) approach is presently being carried out.
A Generalized Eikonal Approximation has been used to deal with  the
FSI effects  and a distorted spin-dependent spectral function, still non relativistic, has been defined.
It has been shown that the formula (\ref{extrac}) can be safely used to obtain  both the Collins and Sivers neutron asymmetries from  the measured Collins and Sivers asymmetries off $^3He$.

Preliminary promising results  (in IA) of the relativistic effects in SIDIS processes were obtained through an evaluation of the LF effective polarizations.
The next step will be the introduction of the FSI  through the GEA within the LFHD.
%
%




\end{document}